\documentclass{article}
\usepackage{spconf,amsmath,graphicx,hyperref}

\usepackage{amssymb}
\usepackage{float}
\usepackage{pifont}
\usepackage{stfloats}
\usepackage[table]{xcolor}

\usepackage{graphicx}
\usepackage{booktabs}

\usepackage{url}

\usepackage{multirow}
\usepackage{caption}
\usepackage{subcaption}

\usepackage{setspace}

\title{Hyperbolic Additive Margin Softmax with Hierarchical Information for Speaker Verification}

\name{Zhihua Fang$^{1,2}$ \qquad Liang He$^{1,2,3,4,\dagger}$
\thanks{
$^\dagger$Corresponding author (heliang@mail.tsinghua.edu.cn).
}}

\address{
$^{1}$School of Computer Science and Technology, Xinjiang University, Urumqi 830017, China\\
$^{2}$Xinjiang Multimodal Information Technology Engineering Research Center, Urumqi 830017, China\\
$^{3}$School of Intelligence Science and Technology, Xinjiang University, Urumqi 830017, China\\
$^{4}$Department of Electronic Engineering, Tsinghua University, Beijing 100084, China
}

\begin{document}
\ninept
\maketitle
\begin{abstract}
Speaker embedding learning based on Euclidean space has achieved significant progress, but it is still insufficient in modeling hierarchical information within speaker features. Hyperbolic space, with its negative curvature geometric properties, can efficiently represent hierarchical information within a finite volume, making it more suitable for the feature distribution of speaker embeddings. In this paper, we propose Hyperbolic Softmax (H-Softmax) and Hyperbolic Additive Margin Softmax (HAM-Softmax) based on hyperbolic space. H-Softmax incorporates hierarchical information into speaker embeddings by projecting embeddings and speaker centers into hyperbolic space and computing hyperbolic distances. HAM-Softmax further enhances inter-class separability by introducing margin constraint on this basis. Experimental results show that H-Softmax and HAM-Softmax achieve average relative EER reductions of 27.84\% and 14.23\% compared with standard Softmax and AM-Softmax, respectively, demonstrating that the proposed methods effectively improve speaker verification performance and at the same time preserve the capability of hierarchical structure modeling. The code will be released at \textcolor[rgb]{0.874, 0.0, 0.486}{\url{https://github.com/PunkMale/HAM-Softmax}}.
\end{abstract}
\section{Introduction}
Speaker verification (SV) aims to determine whether a given speech segment belongs to a target speaker~\cite{review_speaker_modeling2024taslp}. With the development of deep learning, speaker embedding learning methods based on neural networks have gradually replaced traditional approaches and have significantly improved SV performance by learning discriminative speaker embeddings~\cite{x_vectors2018icassp, MVSE2024icassp}. In recent years, researchers have conducted extensive explorations on how to enhance the discriminability of speaker embeddings. Among them, margin-based Softmax loss functions (such as AM-Softmax~\cite{additive_margin2018spl} and AAM-Softmax~\cite{arcface2022pami}) introduce margin constraints in the feature space, effectively reducing intra-class variation and enlarging inter-class distances. At the same time, metric learning and contrastive learning paradigms have also been introduced to alleviate the inconsistency between training and evaluation and further enhance the discriminability of embeddings.

\begin{figure}[tbp]
    \centering
    \includegraphics[width=\linewidth]{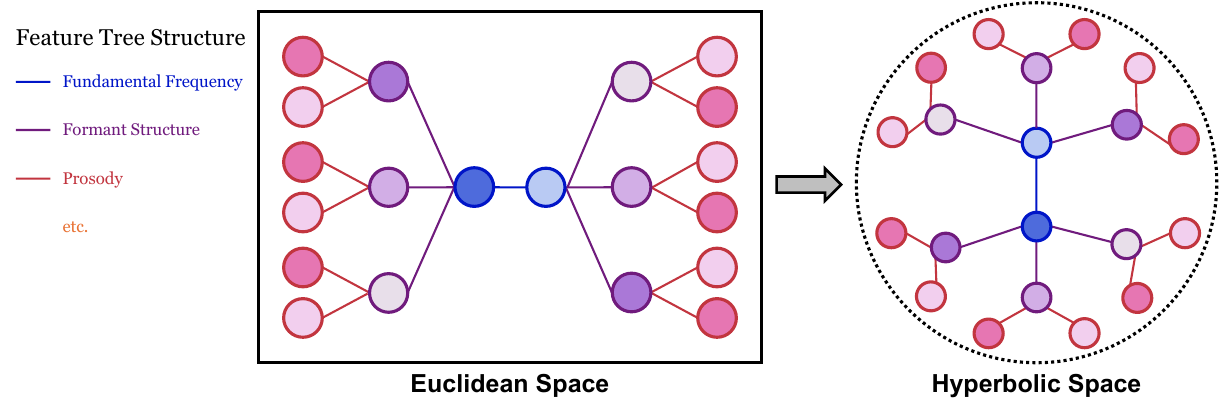}
    \caption{
    \textbf{Illustration of the hierarchical structure of speaker features and hyperbolic space.}
    Left: speaker features contain hierarchical information (e.g., fundamental frequency, formant structure, and prosody, etc.). Right: hyperbolic space with negative curvature geometry is more suitable for representing such hierarchical structures.
    }
    \label{fig:hyper_tree}
\end{figure}

Most existing mainstream methods are modeled in Euclidean space, while speaker features in the real world often contain hierarchical information that exhibits tree-like structures~\cite{decision_tree_speaker2010taslp, BA_LR_attributes2025interspeech} (such as fundamental frequency, formant structures, and prosody; see Fig.~\ref{fig:hyper_tree} left). Euclidean space is insufficient in expressive capacity to fully capture such complex distributions, which limits the separability and generalization ability of speaker embeddings~\cite{hyper_speaker2022spl}. By contrast, thanks to its negative curvature geometry, hyperbolic space can accommodate an exponentially growing number of data points within a finite volume (see Fig.~\ref{fig:hyper_tree} right), making it more suitable for modeling hierarchical structures~\cite{hyper_repre_learn2023icml}. Existing studies have shown that hyperbolic space demonstrates stronger capability in modeling hierarchical structures in text, image, and speech tasks, providing new insights for speaker verification~\cite{hyper_emb2017nips,hyper_img_emb2020cvpr,hyper_speech_separation2024icassp}.

In this paper, we propose the Hyperbolic Softmax (\textbf{H-Softmax}) loss and the Hyperbolic Additive Margin Softmax (\textbf{HAM-Softmax}) loss based on hyperbolic space. H-Softmax incorporates hierarchical information into speaker embedding learning by projecting speaker embeddings into hyperbolic space and introducing hyperbolic distance metrics. On this basis, HAM-Softmax further adds a margin constraint to explicitly enlarge inter-class margins, forcing the model to learn more discriminative speaker representations.
Experimental results show that our methods can effectively improve the performance of speaker verification tasks while preserving the capability of modeling hierarchical structures.

\section{Related Works}
\subsection{Discriminative Loss for Speaker Recognition}
Early speaker verification adopted the standard Softmax with cross-entropy loss, which could effectively classify speakers within the training set but performed poorly in open-set scenarios. To enlarge inter-class distance and reduce intra-class distance, Additive Margin Softmax (AM-Softmax)~\cite{additive_margin2018spl} and Additive Angular Margin Softmax (AAM-Softmax)~\cite{arcface2022pami} were introduced into SV tasks. They impose a penalty on the cosine similarity or angle between embeddings and target class centers, aiming to compress intra-class variation and enlarge inter-class distance so that the model learns discriminative speaker embeddings. Subsequently, various margin-based loss functions~\cite{large_margin_sv2019interspeech,ensemble_amsoftmax2019icassp,dynamic_margin_sv2020interspeech,adaptive_margin_sv2021interspeech} were proposed; however, some studies pointed out that AM-Softmax does not truly implement maximum-margin training and introduced Real AM-Softmax (RAM-Softmax)~\cite{real_amsoftmax2022icassp}, which incorporates a real margin function into softmax training. Another line of research turned to metric learning paradigms, leveraging additive angular margin supervised learning and contrastive learning to enhance speaker separability~\cite{caamcomsup2023icassp}. To mitigate the gap between training and evaluation, Han et al.~\cite{sphereface2_sv2023icassp} introduce the SphereFace2~\cite{SphereFace2_2022iclr} framework with multiple binary classifiers to train speaker models in a pairwise manner, combined with large-margin fine-tuning~\cite{LMFT2023icassp} strategies. In addition, to address practical issues such as speech quality and background noise, the Noise-Aware Quality loss (NAQ-Loss)~\cite{NAQ_loss2024cikm} estimates a quality-aware weight for each utterance and integrates it into the loss function, guiding the model to learn clearer speaker representations.

\subsection{Hyperbolic Geometry in Deep Learning}
In recent years, exploiting the negative curvature property of hyperbolic space to model hierarchical structures in data has attracted increasing interest from researchers~\cite{hyper_repre_learn2023icml}. Nickel and Kiela first demonstrated that Poincaré embeddings significantly outperform Euclidean embeddings in terms of both representational capacity and generalization when applied to data with latent hierarchical structures~\cite{hyper_emb2017nips}. Subsequently, Ganea et al. constructed a complete hyperbolic neural network framework~\cite{hyper_nn2018nips}, laying the foundation for hyperbolic deep learning. Building on this, Khrulkov et al. introduced hyperbolic embeddings into computer vision~\cite{hyper_img_emb2020cvpr}, using hyperbolic space to capture hierarchical semantic relationships among images. Sinha et al. proposed a hyperbolic structural regularization method to accurately embed label hierarchies into learned representations~\cite{hyp_structure2024nips}. In the speech processing domain, Lee et al. proposed a lightweight speaker model based on the Poincaré ball, leveraging hyperbolic geometry to enhance the discriminability and robustness of speaker embeddings~\cite{hyper_speaker2022spl}. Petermann and Kim also applied the Poincaré ball model to the speech separation task, effectively revealing the inherent hierarchical structures in complex speaker mixtures~\cite{hyper_speech_separation2024icassp}.

\section{Methodology}
\noindent \textbf{Motivation.}
In speaker verification tasks, it is generally expected that the model learns discriminative speaker embeddings. Current popular loss functions are usually constructed in Euclidean space based on margin-based softmax. However, the differences between speaker identities in the real world resemble a tree-like structure rather than a flat Euclidean distribution, which makes Euclidean-based loss functions insufficient in modeling complex speaker feature distributions. In contrast, hyperbolic space is naturally suitable for representing hierarchical structures~\cite{hyper_emb2017nips,hyp_structure2024nips}, and its negatively curved geometry allows accommodating an exponential number of points within a finite space. Therefore, we introduce hyperbolic space into speaker embedding learning to enhance inter-class separability and intra-class compactness.

\subsection{Preliminaries}
Among the many equivalent models of hyperbolic space, we adopt the Poincaré ball model to represent hyperbolic space due to its simple metric form and ease of implementation in neural networks~\cite{hyper_emb2017nips}. Let the curvature be $-c$, then the $d$-dimensional Poincaré ball is defined as:
\begin{equation}
    \mathbb{D}_c^d = \{ x \in \mathbb{R}^d : c\|x\|^2 < 1 \},
\end{equation}
where $c>0$ denotes the strength of negative curvature in hyperbolic space, and $\mathbb{D}^d_c$ is the open set inside the unit ball, with boundary points corresponding to infinity. Spaces with different curvature values exhibit different characteristics:
\begin{itemize}
    \item \textit{Small curvature} (e.g., $c=0.01$): approximately Euclidean space, relatively flat and without hierarchical structures.
    \item \textit{Moderate curvature} (e.g., $c=1$): the standard Poincaré Ball, suitable for hierarchical structures.
    \item \textit{Large curvature} (e.g., $c=10$): the distance between points in the space grows faster, enabling better modeling of hierarchical structures but leading to higher numerical instability.
\end{itemize}

In this model, the hyperbolic distance between any two points $x, y \in \mathbb{D}^d_c$ is:
\begin{equation}
    \text{dist}_{\mathbb{D}_c}(x, y) = \text{arcosh} \left( 1 + \frac{2c \|x - y\|^2}{(1 - c\|x\|^2)(1 - c\|y\|^2)} \right),
\end{equation}
where $\text{arcosh} (z)=\ln(z+\sqrt{z^2-1}),\ z\ge 1$ denotes the inverse hyperbolic cosine. In implementation, a lower bound clipping is applied to the input to ensure numerical stability. The hyperbolic distance has the following properties: (1) \textit{Non-negativity}: $\text{dist}_{\mathbb{D}_c}(x, y) \ge 0$. (2) \textit{Unboundedness}: when feature points approach the boundary $|x| \to 1/\sqrt{c}$, $\text{dist}_{\mathbb{D}_c}(x, y) \to \infty.$ (3) \textit{Hierarchy}: since the volume of the Poincaré Ball grows exponentially with the radius, the hyperbolic distance can naturally reflect the hierarchical structure of data.

To prevent embedding vectors from exceeding the boundary of the Poincaré Ball, we define the projection function into the Poincaré Ball as:
\begin{equation}
    \text{proj}(x)= x\cdot\min \left( 1, \frac{1-\varepsilon}{\sqrt{c} \cdot \max(\|x\|, \delta)} \right),
\end{equation}
where $\varepsilon>0$ is used to avoid numerical instability, and $\delta>0$ is a norm lower bound to prevent division by zero.

\begin{figure}[tbp]
    \centering
    \includegraphics[width=\linewidth]{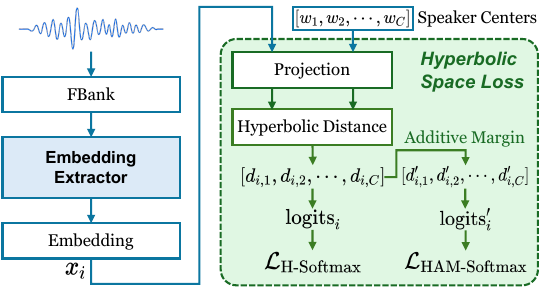}
    \caption{The framework of the proposed method, which incorporates the hyperbolic loss $\mathcal{L}_{\text{H-Softmax}}$ and the hyperbolic additive margin loss $\mathcal{L}_{\text{HAM-Softmax}}$.}
    \label{fig:ham-net}
\end{figure}

\subsection{Hyperbolic Softmax}
Softmax-based loss functions usually use dot product similarity to measure the matching degree between samples and class centers. However, in hyperbolic space, the dot product is no longer suitable as a similarity measure, so we use the negative hyperbolic distance as the similarity metric.

Given an embedding $x_i$ and a class center $w_j$, we first project them onto the Poincaré Ball to obtain $\tilde{x}_i=\text{proj}(x_i)$ and $\tilde{w}_j=\text{proj}(w_j)$, and then compute their hyperbolic distance:
\begin{equation}
    d_{i,j}=\text{dist}_{\mathbb{D}_c}(\tilde{x}_i, \tilde{w}_j).
\end{equation}
We compute hyperbolic distances using the standard Poincaré ball with a fixed curvature of $c=1$. The curvature parameter $c$ is only used to control the effective radius of the Poincaré ball through projection, which implicitly adjusts the strength of geometric constraints imposed on the embeddings.
Then we use the negative hyperbolic distance as logits:
\begin{equation}
    \text{logits}_{i,j} = -s \cdot d_{i,j},
\end{equation}
where $s>0$ denotes the scaling factor. Finally, the Hyperbolic Softmax (H-Softmax) loss is obtained as:
\begin{equation}
    \mathcal{L}_{\text{H-Softmax}} = -\frac{1}{N} \sum_{i=1}^{N} \log \frac{\exp(\text{logits}_{i,y_i})}{\sum_{j=1}^{C} \exp(\text{logits}_{i,j})},
\end{equation}
where $y_i$ denotes the label corresponding to $x_i$, and $C$ denotes the number of classes. By combining classification probabilities with cross-entropy, we obtain the Hyperbolic Softmax (\textbf{H-Softmax}) loss. Compared with the Softmax loss in Euclidean space, H-Softmax introduces hierarchical information into speaker embeddings through hyperbolic space metrics.

\subsection{Hyperbolic Additive Margin Softmax}
To further enhance the inter-class discriminability of speaker embeddings, inspired by margin-based loss functions~\cite{additive_margin2018spl,arcface2022pami}, we introduce an additive margin penalty on top of H-Softmax. By adding a margin to the hyperbolic distance of positive samples, we explicitly enlarge the inter-class margins. In this way, the model not only brings positive sample embeddings closer to their class centers but also needs to overcome the additional margin. We define the adjusted hyperbolic distance as:
\begin{equation}
    d'_{i,j} = 
    \begin{cases} 
    d_{i,j} + m, & j = y_i, \\
    d_{i,j}, & j \neq y_i,
    \end{cases}
\end{equation}
where $m$ denotes the margin, and accordingly $\text{logits}'_{i,j} = -s \cdot d'_{i,j}$. The improved Hyperbolic Additive Margin Softmax (\textbf{HAM-Softmax}) loss is:
\begin{equation}
    \mathcal{L}_{\text{HAM-Softmax}} = -\frac{1}{N} \sum_{i=1}^{N} \log \frac{\exp(\text{logits}'_{i,y_i})}{\sum_{j=1}^{C} \exp(\text{logits}'_{i,j})}.
\end{equation}

With the introduction of the margin, positive samples are encouraged to move closer to their class centers to be correctly classified. Moreover, the margin explicitly increases the penalty for misclassification, encouraging the model to enlarge the separation between different classes in the space, thereby enabling speaker embeddings to not only encode hierarchical information but also become more discriminative.

\section{Experiments}
\subsection{Datasets and Experiments Setup}
\noindent\textbf{Datasets.}
We evaluated the proposed method on VoxCeleb1~\cite{voxceleb1_2017interspeech}, VoxCeleb2~\cite{voxceleb2_2018interspeech}, and CNCeleb~\cite{cnceleb2022specom}. VoxCeleb1 consists of 1,211 speakers with 148,642 utterances, while VoxCeleb2 contains 5,994 speakers and 1,092,009 utterances. For evaluation, we used the clean versions of Vox1-O, Vox1-E, and Vox1-H. CNCeleb comprises over 600,000 utterances from 3,000 Chinese celebrities, with 2,800 speakers in the training set and 200 speakers in the test set. Data augmentation was applied during training using the MUSAN~\cite{musan2015} and RIRs~\cite{rirs2017icassp} corpora.

\noindent\textbf{Training Details.}
For all experiments, the input features were 80-dimensional log-Mel spectrograms. Speaker embeddings were extracted using an ECAPA-TDNN~\cite{ecapa_tdnn2020interspeech} with 1,024 channels, producing 192-dimensional embeddings. The initial learning rate was set to 0.001 and decayed by a factor of 0.97 after each epoch. Optimization was performed using the Adam~\cite{adam2015iclr} optimizer. Training was conducted on a single GeForce RTX 3090 GPU with a batch size of 256. We report performance using equal error rate (EER) and minimum detection cost function (minDCF) with $P_{target}=0.05$. For VoxCeleb1, the model is trained for 150 epochs; for VoxCeleb2 and CNCeleb, the model is trained for 100 epochs.

\noindent\textbf{Baselines.}
For fair comparison, we compare the proposed H-Softmax without margin penalty with the standard Softmax and Softmax with scaling factor $s=30$. For the proposed HAM-Softmax, we compare it with similar margin-based losses, including AM-Softmax~\cite{additive_margin2018spl}, AAM-Softmax~\cite{arcface2022pami}, and Real AM-Softmax~\cite{real_amsoftmax2022icassp}.

\noindent\textbf{Hyperparameters.}
For all loss functions, the margin $m$ is set to 0.2 and the scale $s$ is set to 30, which are settings adopted by most studies. The curvature $c$ of H-Softmax and HAM-Softmax is set to 5 and 3, respectively.

\subsection{Results on VoxCeleb 1\&2 and CNCeleb}
We trained and evaluated models using different loss functions on VoxCeleb1, VoxCeleb2, and CNCeleb datasets, and the results are shown in Table~\ref{table:exp-all}. The results show that the proposed H-Softmax and HAM-Softmax outperform the baselines on multiple test sets. Specifically, H-Softmax shows significant improvements over standard Softmax across all datasets, demonstrating the effectiveness of hyperbolic space in modeling the hierarchical information of speaker embeddings. Notably, even without the margin constraint, H-Softmax still outperforms all margin-based loss functions on CN-Celeb, demonstrating the clear advantage of hyperbolic space in handling complex and cross-domain data. Furthermore, HAM-Softmax, which integrates the margin constraint on top of H-Softmax, achieves the best or second-best results under almost all test conditions. This indicates that the combination of hyperbolic space and the margin mechanism can simultaneously enhance hierarchical structure modeling and inter-class separability, leading to stronger speaker verification performance.

\noindent\textbf{Euclidean-Hyperbolic Additive Margin Softmax.}
We jointly train Real AM-Softmax and HAM-Softmax with a weight ratio of $0.3:0.7$, as reported in Table~\ref{table:exp-all} (`E.-H. AM-Softmax'). This method further improves performance on most test sets, particularly achieving the lowest EER on the Vox-O and Vox-E test sets of VoxCeleb2. This indicates that by relying more on hyperbolic space to model hierarchical structures while incorporating Euclidean space margin constraints to enhance local discriminability, the model is able to demonstrate stronger robustness and generalization ability.

\renewcommand\arraystretch{1.15}
\begin{table*}
    \centering
    \caption{EER (\%) and minDCF results on VoxCeleb1, VoxCeleb2, and CNCeleb, bold indicates the best performance.}
    \label{table:exp-all}
    \resizebox{\textwidth}{!}{
    \begin{tabular}{ccc|cc|cccccc|cc}
        \toprule
        \multicolumn{3}{c}{\textbf{Training data}} \vline & \multicolumn{2}{c}{\textbf{VoxCeleb1}} \vline & \multicolumn{6}{c}{\textbf{VoxCeleb2}} \vline & \multicolumn{2}{c}{\textbf{CNCeleb}} \\
        \hline
        \multicolumn{3}{c}{\textbf{Test Data}} \vline & \multicolumn{2}{c}{\textbf{Vox-O}} \vline & \multicolumn{2}{c}{\textbf{Vox-O}} & \multicolumn{2}{c}{\textbf{Vox-E}} & \multicolumn{2}{c}{\textbf{Vox-H}} \vline & \multicolumn{2}{c}{\textbf{CNCeleb.Eval}} \\
        \cmidrule(lr){1-3} \cmidrule(lr){4-5} \cmidrule(lr){6-7} \cmidrule(lr){8-9} \cmidrule(lr){10-11} \cmidrule(lr){12-13}
        \textbf{Loss Function} & \textbf{Margin} & \textbf{Scale} & \textbf{EER} & \textbf{minDCF} & \textbf{EER} & \textbf{minDCF} &  \textbf{EER} & \textbf{minDCF} & \textbf{EER} & \textbf{minDCF} & \textbf{EER} & \textbf{minDCF}  \\
        \hline
        \textbf{Softmax}& \ding{55} & \ding{55} & 3.318 & 0.236 & 2.074 & 0.146 & 2.322 & 0.161 & 4.321 & 0.255 & 11.676& 0.489 \\
        \textbf{Softmax}& \ding{55} & \ding{51} & 4.702 & 0.317 & 3.046 & 0.217 & 3.180 & 0.214 & 6.082 & 0.353 & 13.765& 0.561 \\
        \textbf{H-Softmax (Ours)} & \ding{55} & \ding{51} & \textbf{2.525} & \textbf{0.192} & \textbf{1.468} & \textbf{0.111} & \textbf{1.507} & \textbf{0.100} & \textbf{2.889} & \textbf{0.180} & \textbf{9.592} & \textbf{0.432} \\
        \hline
        \textbf{AM-Softmax}~\cite{additive_margin2018spl} & \ding{51} & \ding{51} & 2.484 & \textbf{0.176} & 1.298 & 0.102 & 1.461 & 0.097 & 2.732 & 0.168 & 10.865& 0.436 \\
        \textbf{AAM-Softmax}~\cite{arcface2022pami} & \ding{51} & \ding{51} & 2.473 & 0.180 & 1.249 & 0.092 & 1.371 & 0.096 & 2.662 & 0.163 & 10.848& 0.434 \\
        \textbf{Real AM-Softmax}~\cite{real_amsoftmax2022icassp} & \ding{51} & \ding{51} & 2.484 & 0.179 & 1.085 & 0.083 & 1.223 & \textbf{0.080} & \textbf{2.258} & \textbf{0.137} & 10.138& 0.418 \\
        \textbf{HAM-Softmax (Ours)}& \ding{51} & \ding{51} & \textbf{2.409} & 0.186 & \textbf{1.048} & \textbf{0.072} & \textbf{1.178} & \textbf{0.080} & 2.268 & 0.143 & \textbf{9.498} & \textbf{0.413}\\
        \hline
        \textbf{E.-H. AM-Softmax (Ours)}& \ding{51} & \ding{51} & 2.430 & 0.177 & \textbf{1.005} & 0.077 & 1.184 & \textbf{0.078} & \textbf{2.239} & 0.139 & 9.778 & 0.420\\
        \bottomrule
    \end{tabular}}
\end{table*}

\subsection{Ablation Study}
\begin{figure}[tbp]
    \centering
    \includegraphics[width=\linewidth]{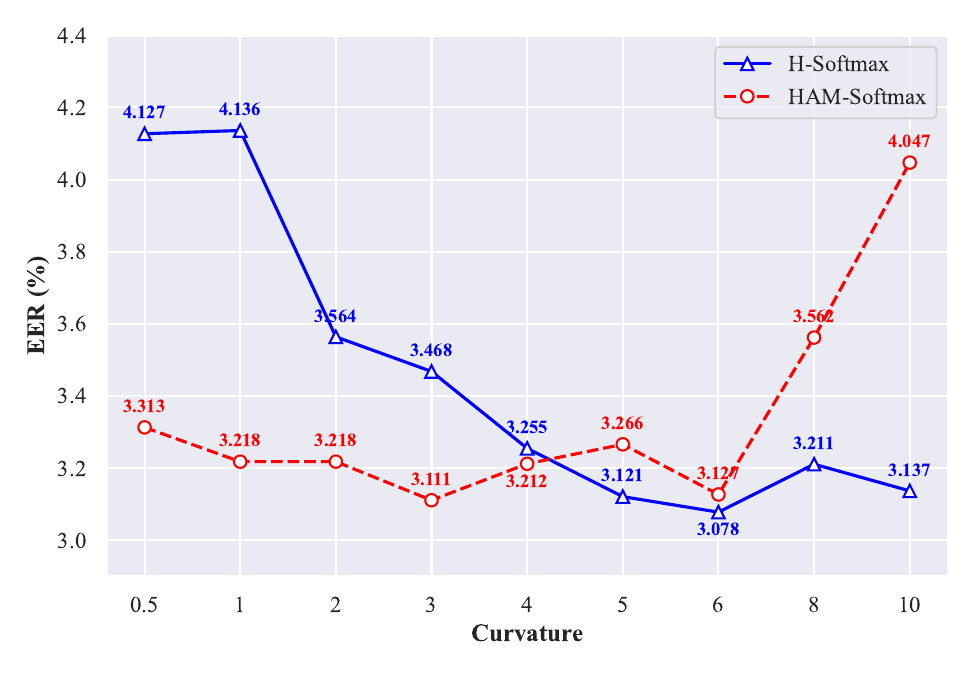}
    \caption{EER (\%) results of the curvature ablation study on VoxCeleb1 (without data augmentation).}
    \label{fig:abla_curvature}
\end{figure}

\noindent\textbf{Ablation analysis of curvature.}
We study the influence of different curvature values on H-Softmax and HAM-Softmax on VoxCeleb1. This experiment is conducted without data augmentation to evaluate the pure effect of curvature values on model performance. The results are shown in Fig.~\ref{fig:abla_curvature}. Overall, it can be observed that in the small curvature range, the performance of HAM-Softmax is significantly better than H-Softmax; in the medium curvature range, the difference between the two is small with performance fluctuations; and when the curvature further increases, the EER of HAM-Softmax increases significantly, while H-Softmax still maintains a low level.

According to the experimental observations, when the curvature parameter is small, the Poincaré ball has a larger effective radius and imposes weak geometric constraints on the embeddings, resulting in behavior close to Euclidean space. In this case, H-Softmax is insufficient to enlarge inter-class separation based solely on hyperbolic distance, leading to limited discriminability, whereas HAM-Softmax explicitly improves class separability by introducing an additive margin.
As the curvature parameter increases, the effective radius of the Poincaré ball decreases and stronger geometric constraints are imposed, naturally amplifying inter-class differences and enabling H-Softmax to achieve strong discriminative performance. However, applying a fixed additive margin under such conditions may cause over-separation and introduce numerical or optimization instability, which degrades the performance of HAM-Softmax.

\noindent\textbf{Discussion.}
Previous work has explored the application of hyperbolic space combined with triplet loss in speaker recognition~\cite{hyper_speaker2022spl}, but it only reported results for low curvatures ($c=[0.3, 0.03, 0.003]$). In contrast, we combined hyperbolic space with Softmax to develop two classification losses, and we found better performance at higher curvatures ($3 \le c \le 6$), confirming that high-curvature hyperbolic space better models hierarchical structures. Additionally, our two proposed loss functions can serve as plug-and-play modules for other frameworks.

\begin{table}[htbp]
    \centering
    \caption{EER (\%) and minDCF results of H-Softmax and HAM-Softmax with different scales and margins on VoxCeleb1 (without data augmentation).}
    \label{table:abla_scale_margin}
    \resizebox{\linewidth}{!}{%
    \begin{tabular}{c|ccc|ccccc}
    \toprule
    \textbf{Loss} & \multicolumn{3}{c}{\textbf{H-Softmax}} \vline & \multicolumn{5}{c}{\textbf{HAM-Softmax}} \\
    \cmidrule(r){1-1} \cmidrule(lr){2-4} \cmidrule(l){5-9}
    \textbf{Parameter} & $s=1$ & $s=30$ & $s=60$ & $m=0.05$ & $m=0.1$ & $m=0.2$ & $m=0.3$ & $m=0.4$ \\
    \midrule
    \textbf{EER}    & 16.710 & 3.121 & 4.228 & 3.250 & 3.116 & 3.111 & 3.243 & 3.415 \\
    \textbf{minDCF} & 0.976  & 0.221 & 0.260 & 0.221 & 0.228 & 0.238 & 0.248 & 0.250 \\
    \bottomrule
    \end{tabular}%
    }
\end{table}

\noindent\textbf{Ablation analysis of margin and scale.}
As shown in Table~\ref{table:abla_scale_margin}, the scaling factor has a significant impact on the performance of H-Softmax; an appropriate $s$ can effectively enhance discriminative capability. In contrast, HAM-Softmax exhibits more stable performance across different margins. Introducing an appropriate margin can maintain inter-class separability while preserving good training stability. However, when the margin is further increased to 0.3 or 0.4, both EER and minDCF show an upward trend, indicating that an excessively large margin leads to over-separation and degrades model performance.

\section{Conclusions}
This paper proposes H-Softmax and HAM-Softmax based on hyperbolic space, aimed at enhancing hierarchical modeling and inter-class separability of speaker embeddings. Experimental results show that H-Softmax outperforms margin-based methods on cross-domain complex data, while HAM-Softmax achieves the best or second-best performance on all datasets, further validating the effectiveness of margin constraints in hyperbolic space. Through extensive ablation studies, we reveal the effects of curvature, scale, and margin on model performance, indicating that reasonable parameter settings can fully exploit the advantages of hyperbolic space modeling and provide effective insights for speaker embedding learning.

\section{Acknowledgements}
This work was supported in part by the National Natural Science Foundation of China under Grant 62366051, in part by the State Grid Xinjiang Electric Power Company and Xinjiang Siji Information Technology Co., Ltd. under Grant SGITXX00ZHXX2200262, and in part by the ``Small Group'' Aid Xinjiang Project under Grant 51052501207.

\bibliographystyle{IEEEbib}
\bibliography{reference}

\end{document}